\documentclass[reprint,superscriptaddress,aps,pra]{revtex4-2}
\usepackage{amsmath}
\usepackage{graphicx}
\usepackage{subfigure}
\usepackage{amssymb}
\usepackage{hyperref}
\hypersetup{colorlinks=true, citecolor=blue, urlcolor=blue, linkcolor=blue}

\begin{document}
	\title{Modifying $n$-qubit controlled-$ZX$ gate to be $n$-qubit Toffoli gate}
	
	\author{Jian Leng}
	\affiliation{ State Key Laboratory of Low Dimensional Quantum Physics, Department of Physics, \\ Tsinghua University, Beijing 100084, China}
	\author{Fan Yang}
	\affiliation{ State Key Laboratory of Low Dimensional Quantum Physics, Department of Physics, \\ Tsinghua University, Beijing 100084, China}
	\author{Xiang-Bin Wang}
	\email{ xbwang@mail.tsinghua.edu.cn}
	\affiliation{ State Key Laboratory of Low Dimensional Quantum Physics, Department of Physics, \\ Tsinghua University, Beijing 100084, China}
	\affiliation{ Jinan Institute of Quantum technology, SAICT, Jinan 250101, China}
	\affiliation{ Shanghai Branch, CAS Center for Excellence and Synergetic Innovation Center in Quantum Information and Quantum Physics, University of Science and Technology of China, Shanghai 201315, China}
	\affiliation{ Shenzhen Institute for Quantum Science and Engineering, and Physics Department, Southern University of Science and Technology, Shenzhen 518055, China}
	\affiliation{ Frontier Science Center for Quantum Information, Beijing, China}
	
	\begin{abstract}
		The decomposition for controlled-$ZX$ gate in [\href{https://journals.aps.org/pra/abstract/10.1103/PhysRevA.87.062318}{Phys. Rev. A \textbf{87}, 062318 (2013)}] has a shallow circuit depth $8n-20$ with no ancilla. Here we modify this decomposition to decompose $n$-qubit Toffoli gate with only $2n-3$ additional single-qubit gates. The circuit depth is unchanged and no ancilla is needed. We explicitly show that the circuit after decomposition can be easily constructed in present physical systems.
	\end{abstract}
	
	
	\maketitle
	
	\section{Introduction}
	Quantum algorithms \cite{shor1994algorithms,grover1996fast} have excellent advantages on lots of computational tasks for which classical algorithms are powerless. To realize quantum algorithms in practical device, they have to be decomposed into practicable quantum gates that can be constructed in physical systems \cite{bruzewicz2019trapped,arakawa2020progress,gaita2019molecular,chatterjee2021semiconductor,kjaergaard2020superconducting,adams2019rydberg}. Particularly, $n$-qubit Toffoli gate or $n$-qubit controlled-$X$ gate is widely applied in quantum algorithms. How to decompose $n$-qubit Toffoli gate with shallower circuit depth, smaller circuit size, and fewer ancillary qubits is an important problem \cite{PhysRevA.52.3457,liu2008analytic,PhysRevA.93.032318,lanyon2009simplifying,he2017decompositions,PhysRevA.87.062318,PhysRevA.94.026301}. It is at first decomposed with $\mathcal{O}(n^2)$ circuit depth and size \cite{PhysRevA.52.3457,liu2008analytic,PhysRevA.93.032318}, then these costs are reduced to $\mathcal{O}(n)$ with the help of ancillary qubits \cite{lanyon2009simplifying,he2017decompositions}. Without any ancillary qubits, $n$-qubit controlled-$ZX$ gate can be decomposed with $\mathcal{O}(n)$ circuit depth and $\mathcal{O}(n^2)$ circuit size \cite{PhysRevA.87.062318,PhysRevA.94.026301}. Precisely, its circuit depth is very shallow: $8n-20$ \cite{da2022linear}, compared with the common order of magnitude $\sim10^2n$ which is the circuit depth of present decomposition methods with one ancillary qubit. Besides the advantages that shadow depth and no ancilla, the decomposition in Ref. \cite{PhysRevA.87.062318} can be easily constructed in present physical systems as we show in Section IV. However, this remarkable decomposition cannot be directly applied on $n$-qubit controlled-$X$ gate \cite{PhysRevA.94.026301} and we need to modify the method. A trivial way is that executing an $n$-qubit controlled-$Z$ gate before $n$-qubit controlled-$ZX$ gate. This deepens the circuit depth, enlarges the circuit size and probably leads in some ancillary qubits. More efficient way is that modifying the original $n$-qubit controlled-$ZX$ gate. With replacing $2n-3$ specified controlled-$R_x(\pi/k)$ gates by controlled-$\sqrt[k]{X}$ gates ($k=2^1,2^2,...,2^{n-2}$), the new circuit is exactly $n$-qubit controlled-$X$ gate \cite{da2022linear}. This keeps the circuit depth unchanged and no ancilla is needed. But controlled-$\sqrt[k]{X}$ gates are not common-used quantum gates and generally they should be decomposed into one $\theta$-phase gate $S(\theta)=|0\rangle\langle0|+e^{i\theta}|1\rangle\langle1|$, two $R_y(\theta)$ gates, three $R_z(\theta)$ gates and two CNOT gates \cite{PhysRevA.52.3457,Nielsen2002}. This leads to additional seven gates for implementing one controlled-$\sqrt[k]{X}$ gate and the total circuit size grows about $(2n-3)*7=14n-21$ gates. Here, we show that with a very simple modification, the circuit in Ref. \cite{PhysRevA.87.062318} can be changed to realize $n$-qubit Toffoli gate. We add $2n-3$ $\theta$-phase gates into the original circuit and prove that the new circuit is the $n$-qubit Toffoli gate. The circuit depth of our new circuit is the same as the original one and no ancilla is needed.
	
	\section{$n$-qubit Toffoli gate}
	
\begin{figure*}[htbp]
		\begin{minipage}{1.1\linewidth}
			\hspace{-2cm}\includegraphics[width=1\linewidth]{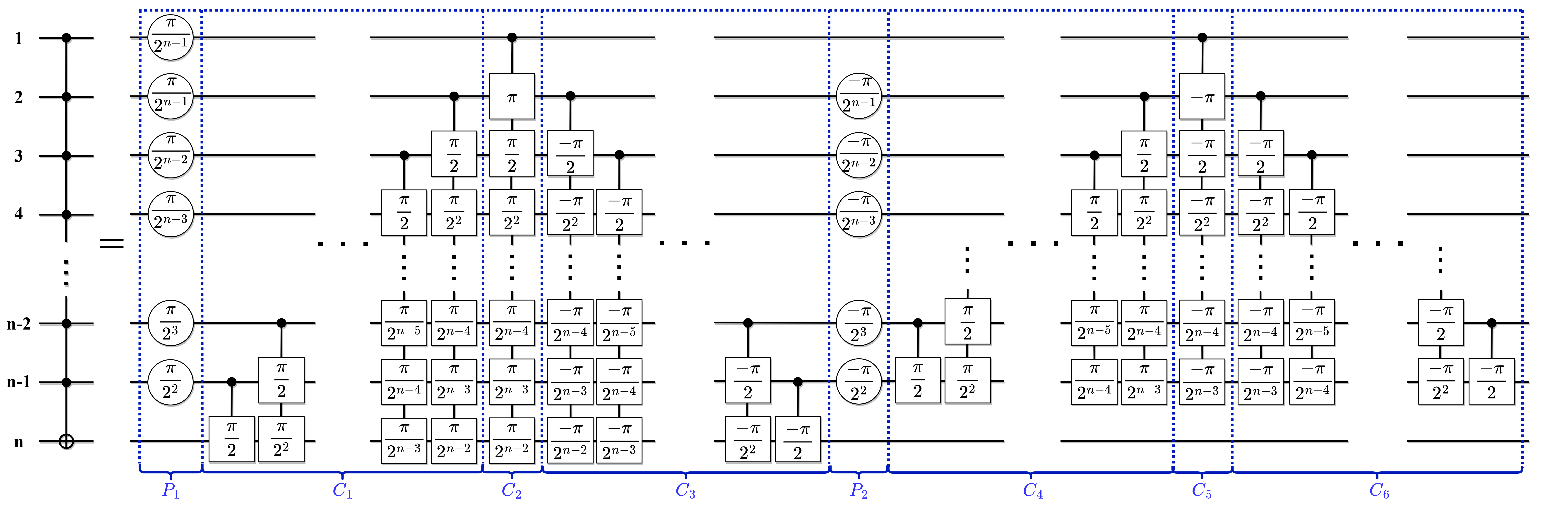}
		\end{minipage}
		\caption{\label{Toffoli_circuit}The decomposition for $n$-qubit Toffoli gate. It is divided into eight parts: $P_1+C_1+C_1+C_3+P_2+C_4+C_5+C_6$. $C_1-C_6$ are the same as $n$-qubit controlled-$ZX$ gate in Ref. \cite{PhysRevA.87.062318} and they only include controlled-$R_x(\theta)$ gates. $P_1$ and $P_2$ are new parts that we add and they include $2n-3$ single-qubit $\theta$-phase gates $S(\theta)=|0\rangle\langle0|+e^{i\theta}|1\rangle\langle1|$.}
	\end{figure*}

	Our circuit for $n$-qubit Toffoli gate is shown in Fig. \ref{Toffoli_circuit}.
	Compared to the original circuit in Ref. \cite{PhysRevA.87.062318}, our circuit only adds $2n-3$ additional $\theta$-phase gates. Obviously, the circuit depth does not grow and no ancilla is needed.
	
	\textbf{Theorem 1:} The circuit in Fig. \ref{Toffoli_circuit} is $n$-qubit Toffoli gate.
	
	\textbf{Proof:} Focus on the last qubit $n$. The first gate applied on qubit $n$ is controlled-$R_x(\frac{\pi}{2})$. We combine this gate with $S(\frac{\pi}{2^2})$ in $P_1$ and obtain:
	\begin{align}
	&\left(
	\begin{array}{cc}
		I & 0\\
		0 & R_x(\frac{\pi}{2})
	\end{array}
	\right)S\left(\frac{\pi}{2^2}\right)\otimes I=
	\left(
	\begin{array}{cc}
		I & 0\\
		0 & e^{-iX\frac{\pi}{2^2}}
	\end{array}
	\right)
	\left(
	\begin{array}{cc}
		I & 0\\
		0 & e^{i\frac{\pi}{2^2}}
	\end{array}
	\right)\nonumber\\
	=&
	\left(
	\begin{array}{cc}
		I & 0\\
		0 & e^{i\frac{\pi}{2^2}}e^{-iX\frac{\pi}{2^2}}
	\end{array}
	\right)
	=
	\left(
	\begin{array}{cc}
		I & 0\\
		0 & \sqrt{X}
	\end{array}
	\right),
	\end{align}
	where $R_x(\theta)=e^{-iX\frac{\theta}{2}}$ has been used and $X$ is the Pauli-X matrix. So the combination is a controlled-$\sqrt{X}$ gate. The second gate applied on qubit $n$ is controlled-$R_x(\frac{\pi}{2^2})$. Similarly, we combine this gate with $S(\frac{\pi}{2^3})$ in $P_1$ and obtain a controlled-$\sqrt[2^2]{X}$ gate. Following the same procedure, we combine each controlled-$R_x(\frac{\pi}{2^t})$ gate applied on qubit $n$ in $C_1$ and $C_2$ with each $S(\frac{\pi}{2^{t+1}})$ in $P_1$ and obtain a controlled-$\sqrt[2^t]{X}$ gate. Also, we combine each controlled-$R_x(\frac{-\pi}{2^t})$ gate applied on qubit $n$ in $C_3$ with each $S(\frac{-\pi}{2^{t+1}})$ in $P_2$ and obtain each controlled-$\sqrt[2^t]{X}^\dagger$ gate. Consequently, $P_1$ and $P_2$ are combined into $C_1-C_3$ and each controlled-$R_x(\frac{\pi}{2^t})$ (or -$R_x(\frac{-\pi}{2^t})$) gate applied on qubit $n$ is replaced by each controlled-$\sqrt[2^t]{X}$ (or -$\sqrt[2^t]{X}^\dagger$) gate. Applying \textit{Theorem 1} in Ref. \cite{da2022linear}, we conclude that the circuit after combination is $n$-qubit Toffoli gate. $\square$
	
	\section{Local gates}
	
	Local gates include single-qubit gates for one qubit and two-qubit gates for two neighboring qubits \cite{PhysRevA.87.062318}. Local gates do not involve coherently operating three or more qubits and also do not need to operate the interaction between two distant qubits. The circuit in Fig. \ref{Toffoli_circuit} can be further decomposed into local gates according to \textit{Theorem 3} in Ref. \cite{PhysRevA.87.062318}. For practical application, here we explicitly show the decomposition for $n$-qubit Toffoli gate with local gates.
	
\begin{figure}[htbp]
		\begin{minipage}{1\linewidth}
			\leftline{\textbf{(a)}~$C_1:$}
			\vspace{0.3cm}
			\includegraphics[width=1\linewidth]{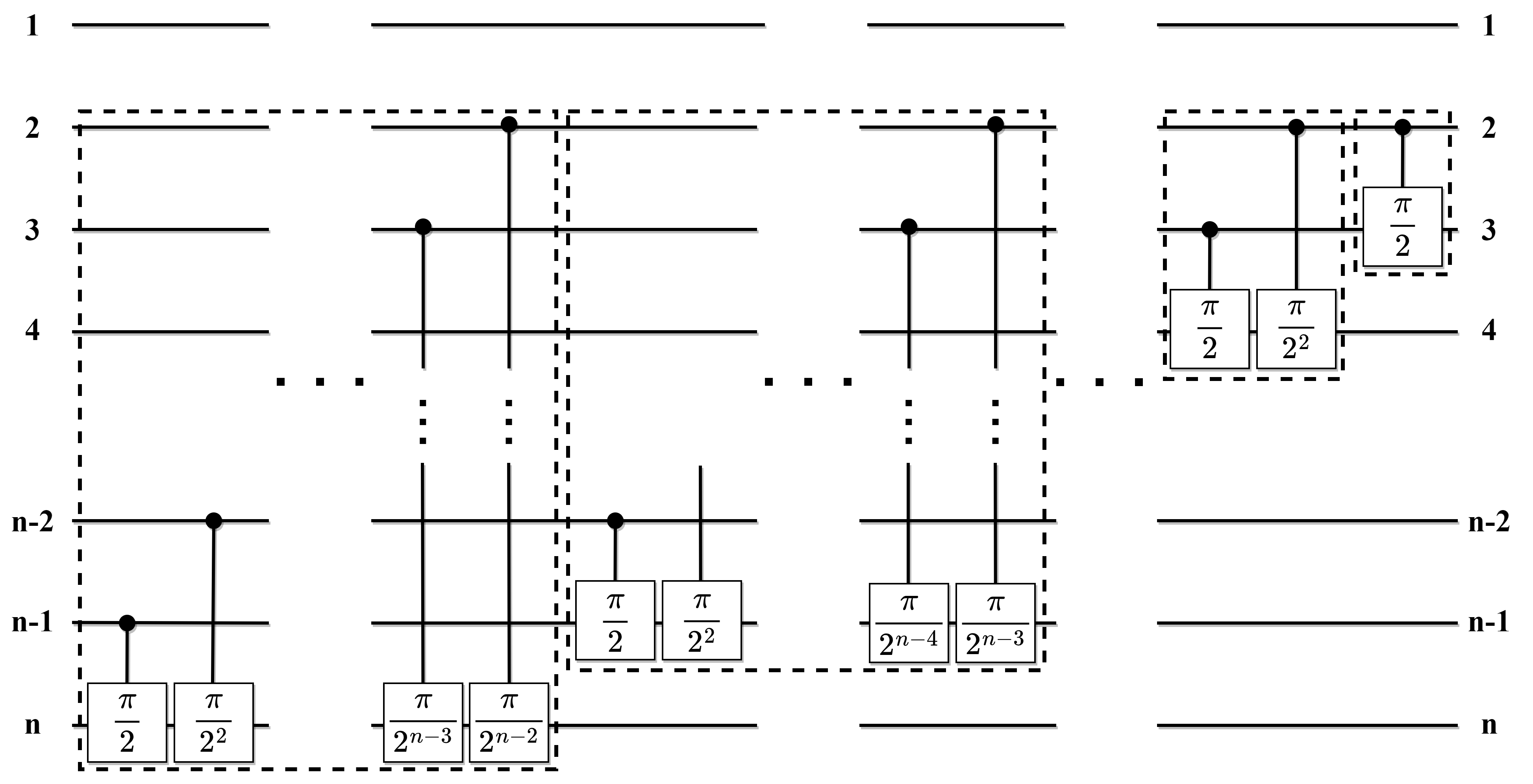}
			\vspace{0.3cm}
		\end{minipage}
		\begin{minipage}{1.05\linewidth}
			\leftline{\textbf{(b)}~$C_1^\prime:$}
			\vspace{0.3cm}
			\includegraphics[width=1\linewidth]{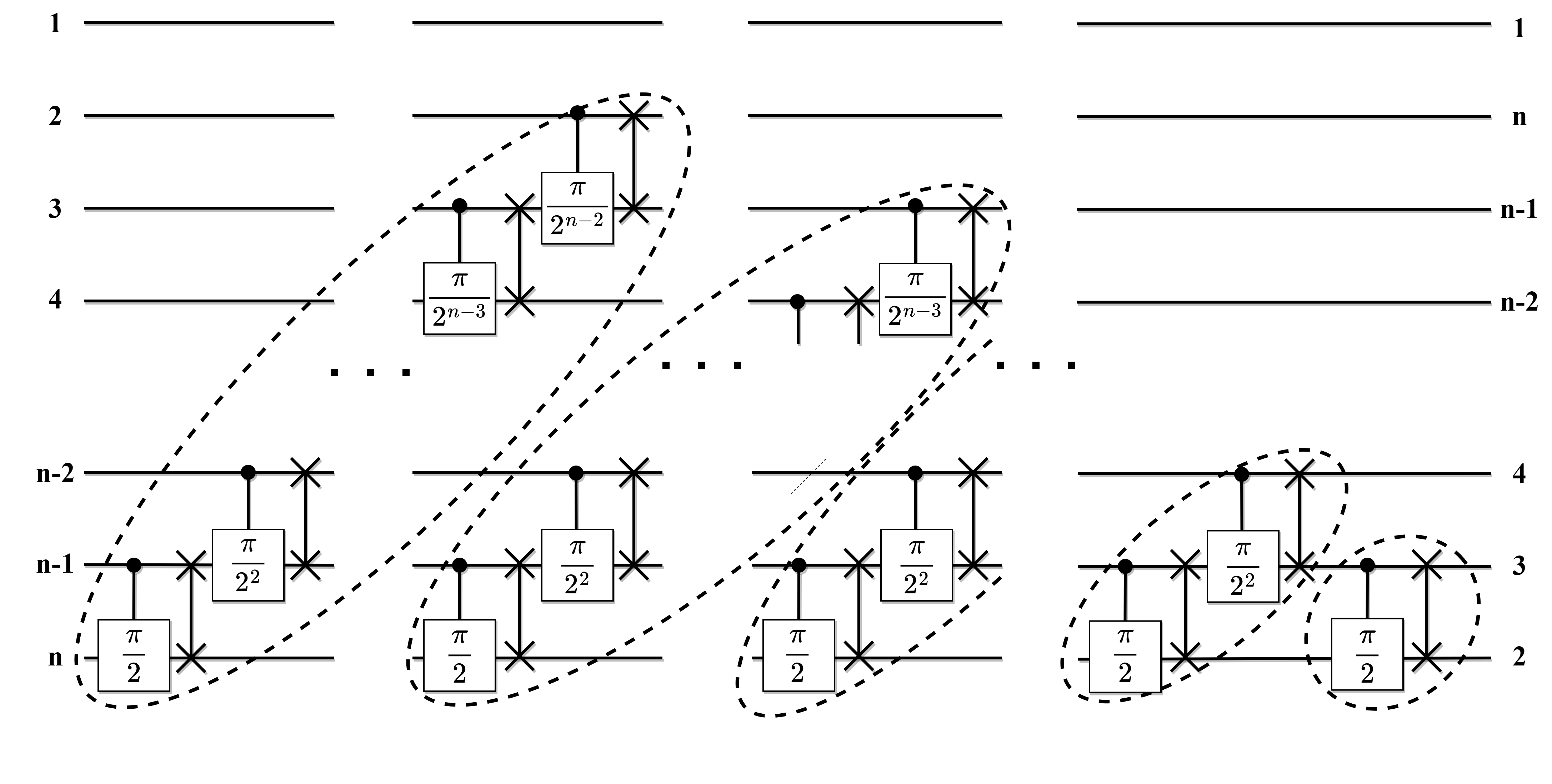}
		\end{minipage}
		\caption{\label{C_1_circuit}(a) An equivalent circuit for $C_1$ that only includes two-qubit controlled-$R_x(\theta)$ gates. It is divided into many blocks (dashed rectangles). (b) An equivalent circuit $C_1^\prime$ for $C_1$ that only includes local controlled-$R_x(\theta)$ and local swap gates. The qubit ordering is changed: $\{1,2,3,...,n-1,n\}\to\{1,n,n-1,...,3,2\}$. Each block (dashed ellipse) corresponds to a block in (a).}
	\end{figure}

\begin{figure}[htbp]
		\begin{minipage}{0.425\linewidth}
			\leftline{\textbf{(a)}~$C_2:$}
			\vspace{0.3cm}
			\includegraphics[width=1\linewidth]{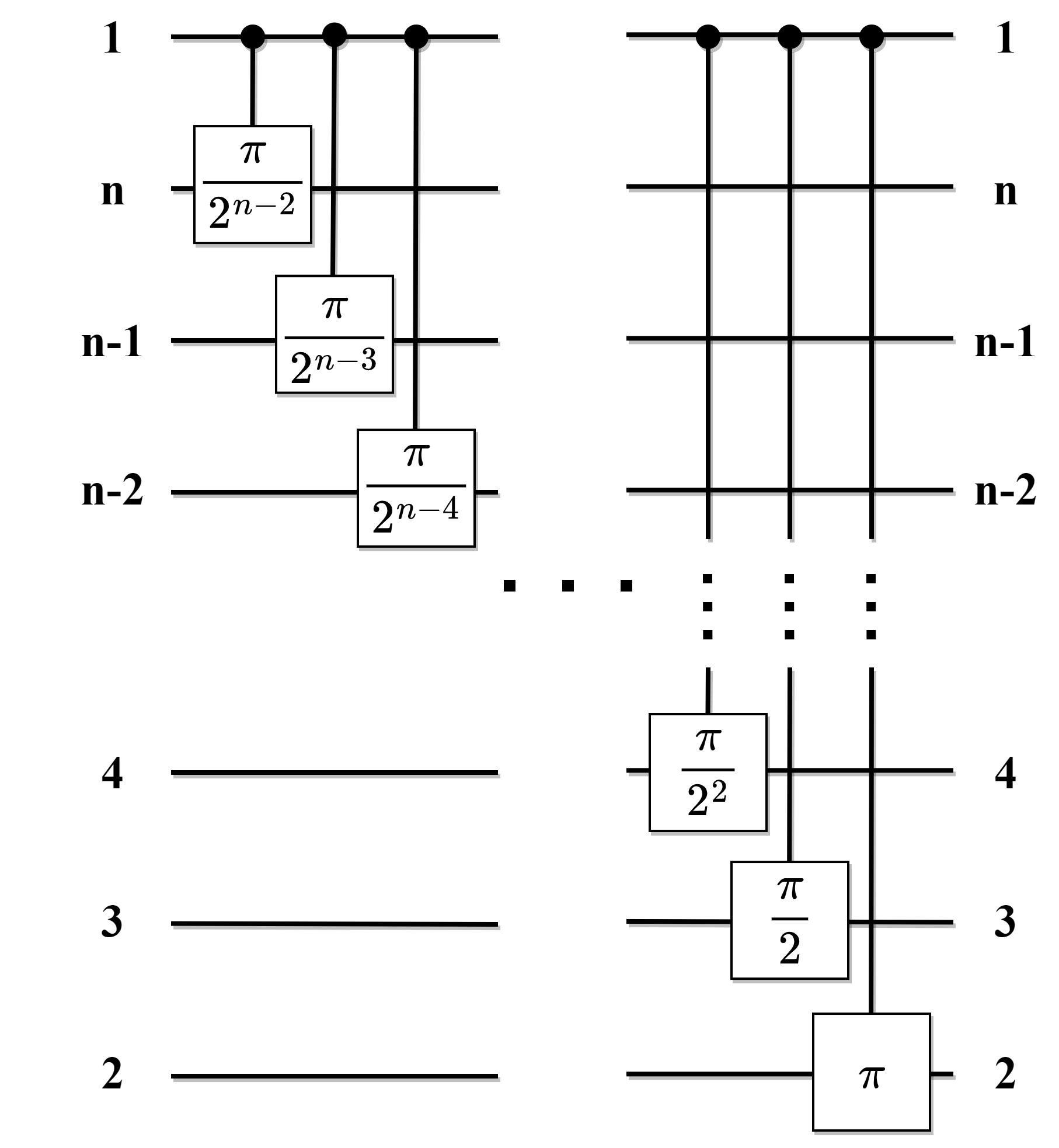}
		\end{minipage}
		\begin{minipage}{0.555\linewidth}
			\leftline{\textbf{(b)}~$C_2^\prime:$}
			\vspace{0.3cm}
			\includegraphics[width=1\linewidth]{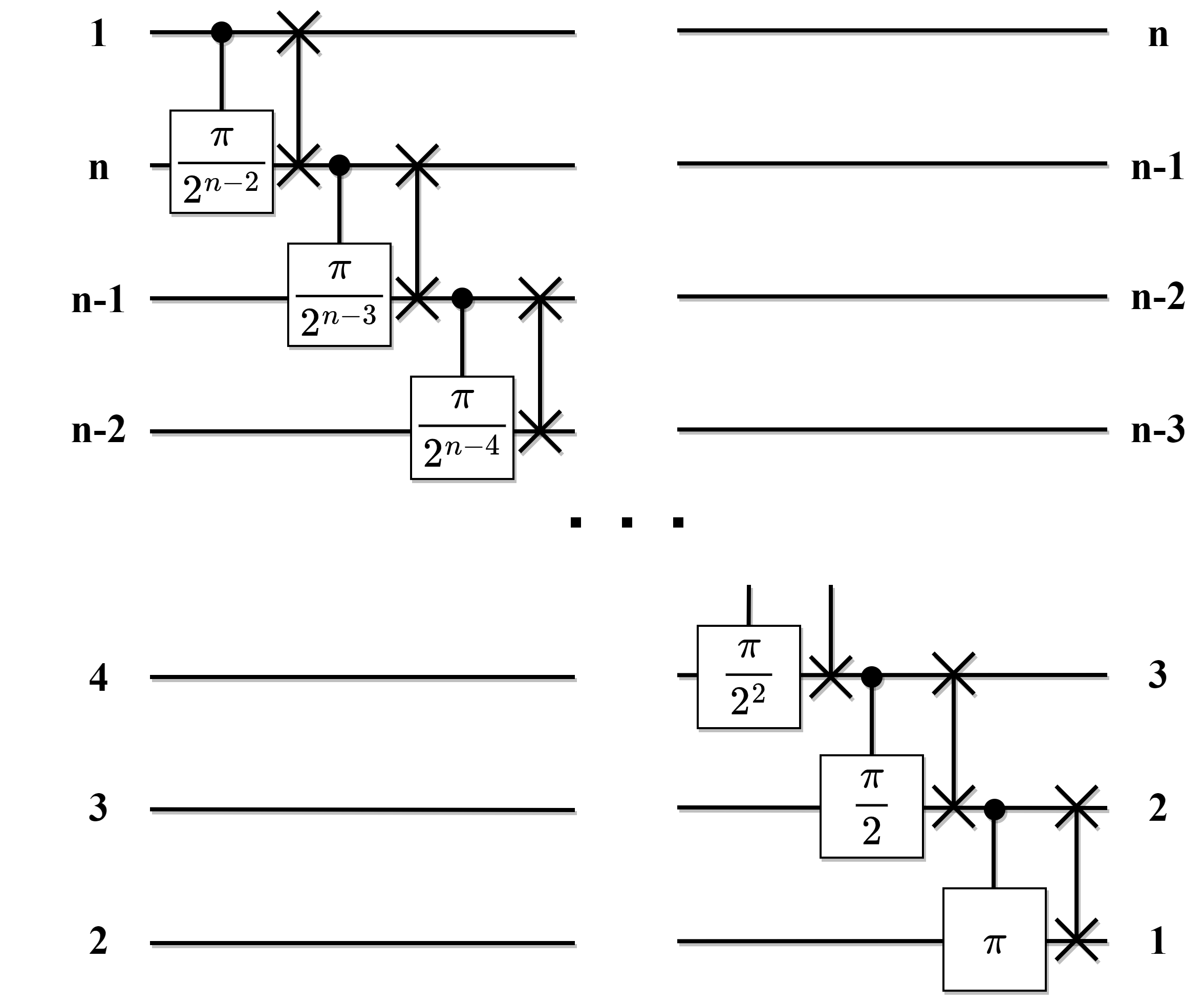}
		\end{minipage}
		\caption{\label{C_2_circuit}(a) An equivalent circuit for $C_2$ that only includes two-qubit controlled-$R_x(\theta)$ gates. The qubit ordering is $\{1,n,n-1,...,3,2\}$. (b) An equivalent circuit $C_2^\prime$ for $C_2$ that only includes local controlled-$R_x(\theta)$ and local swap gates. The qubit ordering $\{1,n,n-1,...,3,2\}$ is changed to $\{n,n-1,n-2,...,2,1\}$.}
	\end{figure}

\begin{figure}[htbp]
		\begin{minipage}{1\linewidth}
			\leftline{\textbf{(a)}~$C_3:$}
			\vspace{0.3cm}
			\includegraphics[width=1\linewidth]{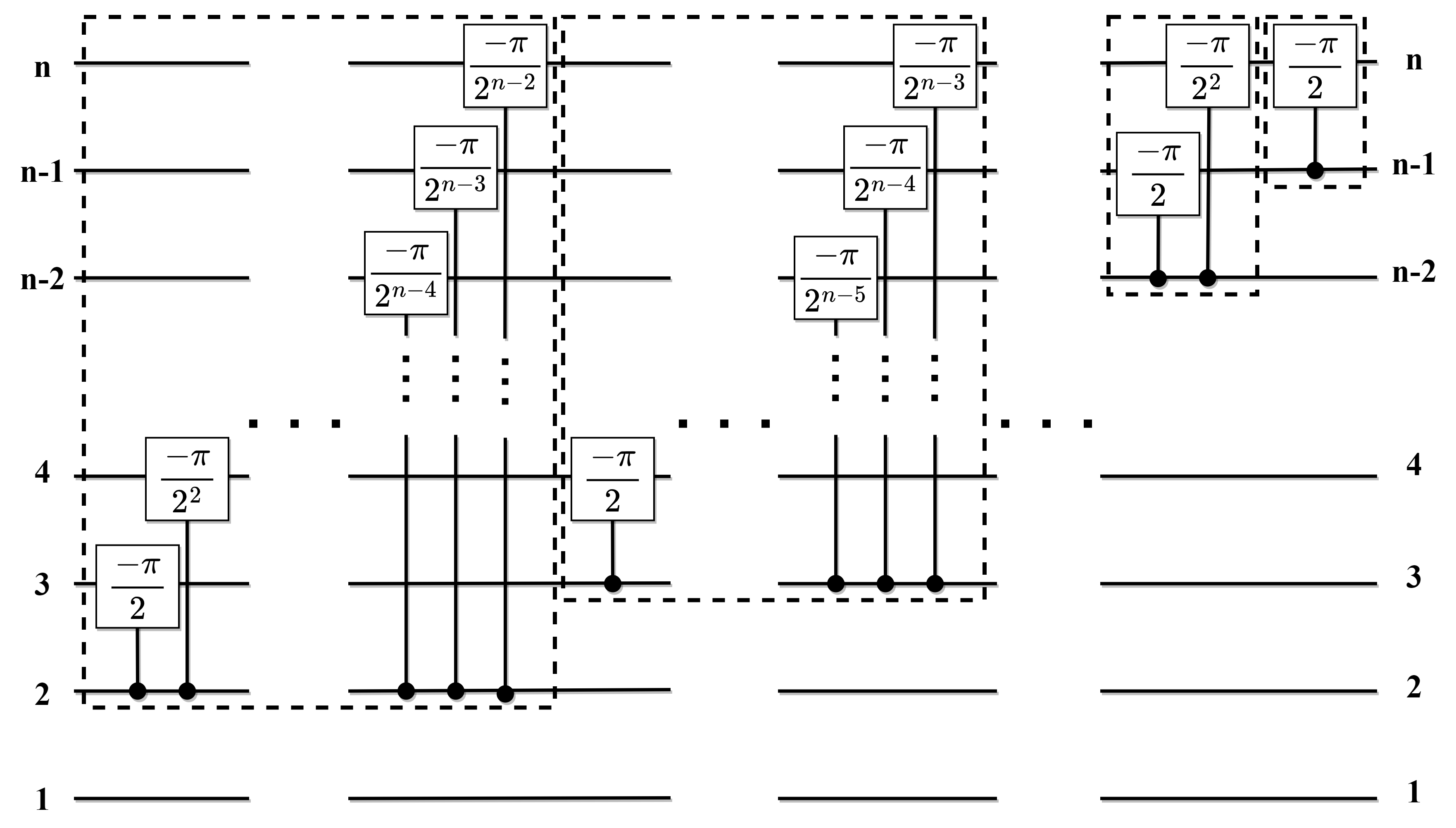}
			\vspace{0.3cm}
		\end{minipage}
		\begin{minipage}{1.05\linewidth}
			\leftline{\textbf{(b)}~$C_3^\prime:$}
			\vspace{-0.3cm}
			\includegraphics[width=1\linewidth]{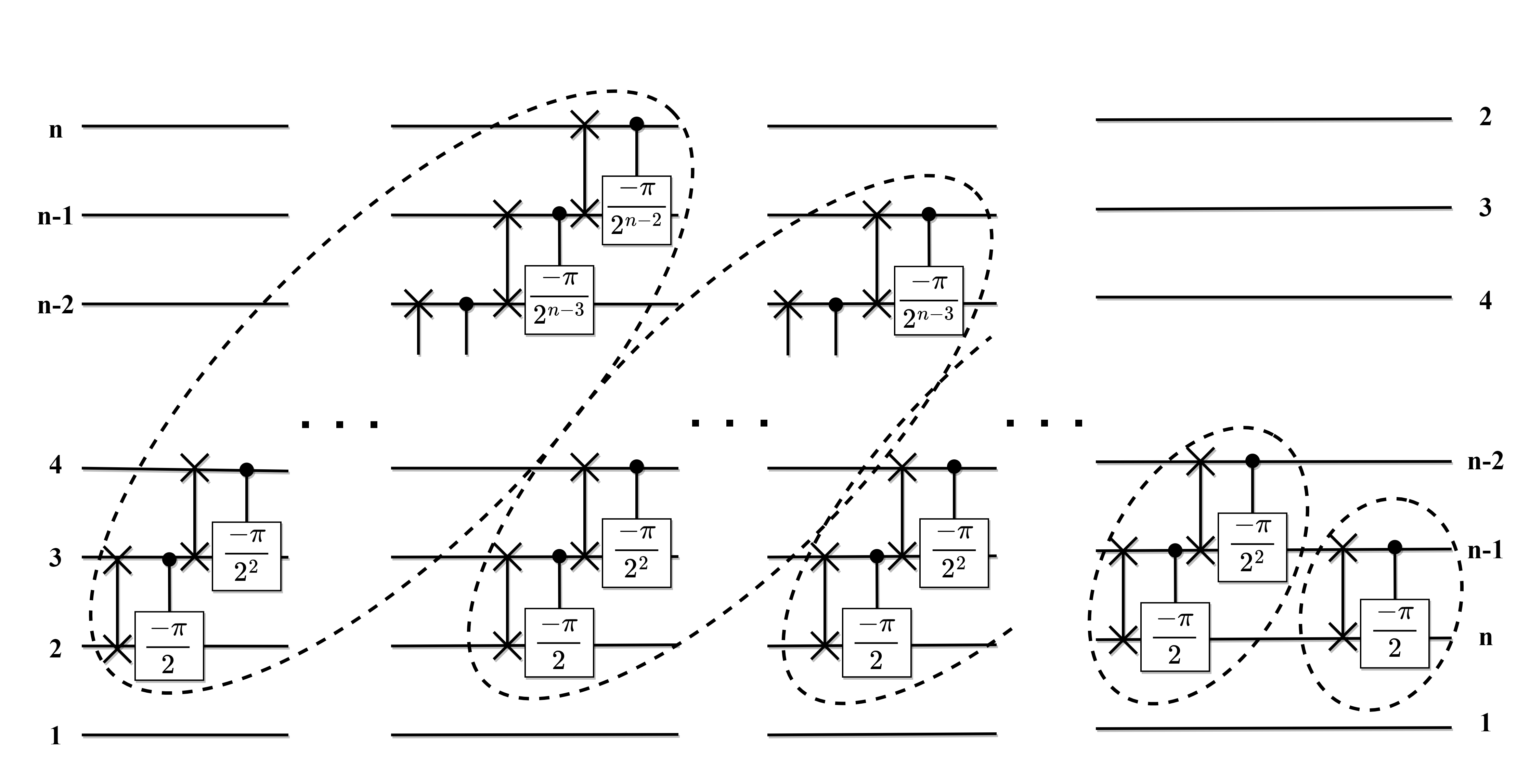}
		\end{minipage}
		\caption{\label{C_3_circuit}(a) An equivalent circuit for $C_3$ that only includes two-qubit controlled-$R_x(\theta)$ gates. It is divided into many blocks (dashed rectangles). (b) An equivalent circuit $C_3^\prime$ for $C_3$ that only includes local controlled-$R_x(\theta)$ and local swap gates. The qubit ordering is changed: $\{n,n-1,n-2,...,2,1\}\to\{2,3,...,n-1,n,1\}$. Each block (dashed ellipse) corresponds to a block in (a).}
	\end{figure}

\begin{figure}[htbp]
		\begin{minipage}{0.43\linewidth}
			\leftline{~$C_{3.5}^\prime:$}
			\vspace{0.3cm}
			\includegraphics[width=1\linewidth]{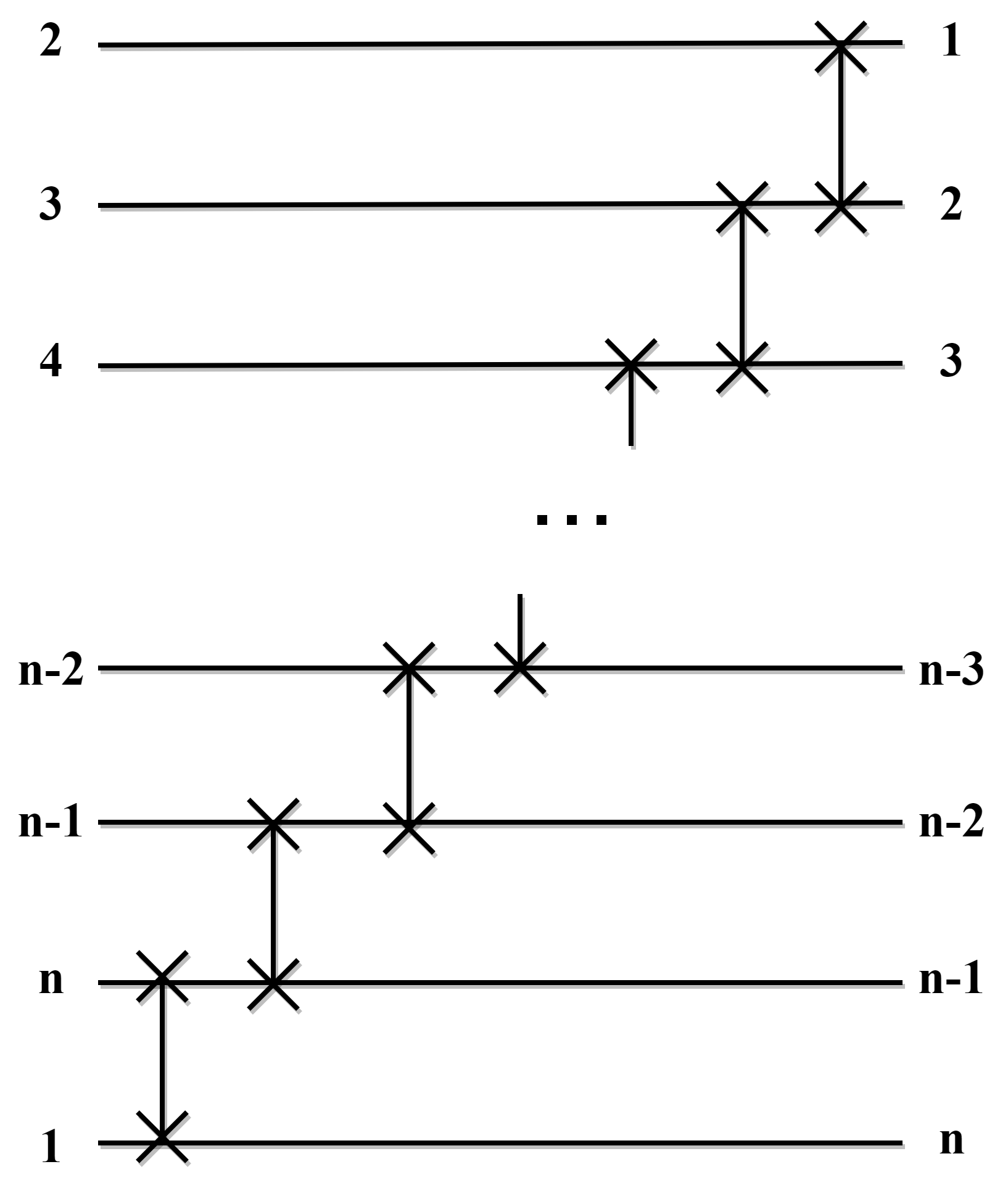}
		\end{minipage}
		\caption{\label{C_3_5}The circuit $C_{3.5}^\prime$ resets the qubit ordering $\{2,3,...,n-1,n,1\}$ to $\{1,2,3,...,n-1,n\}$.}
	\end{figure}

\begin{figure}[htbp]
		\begin{minipage}{1\linewidth}
			\leftline{\textbf{(a)}~$C_4:$}
			\vspace{0.3cm}
			\includegraphics[width=1\linewidth]{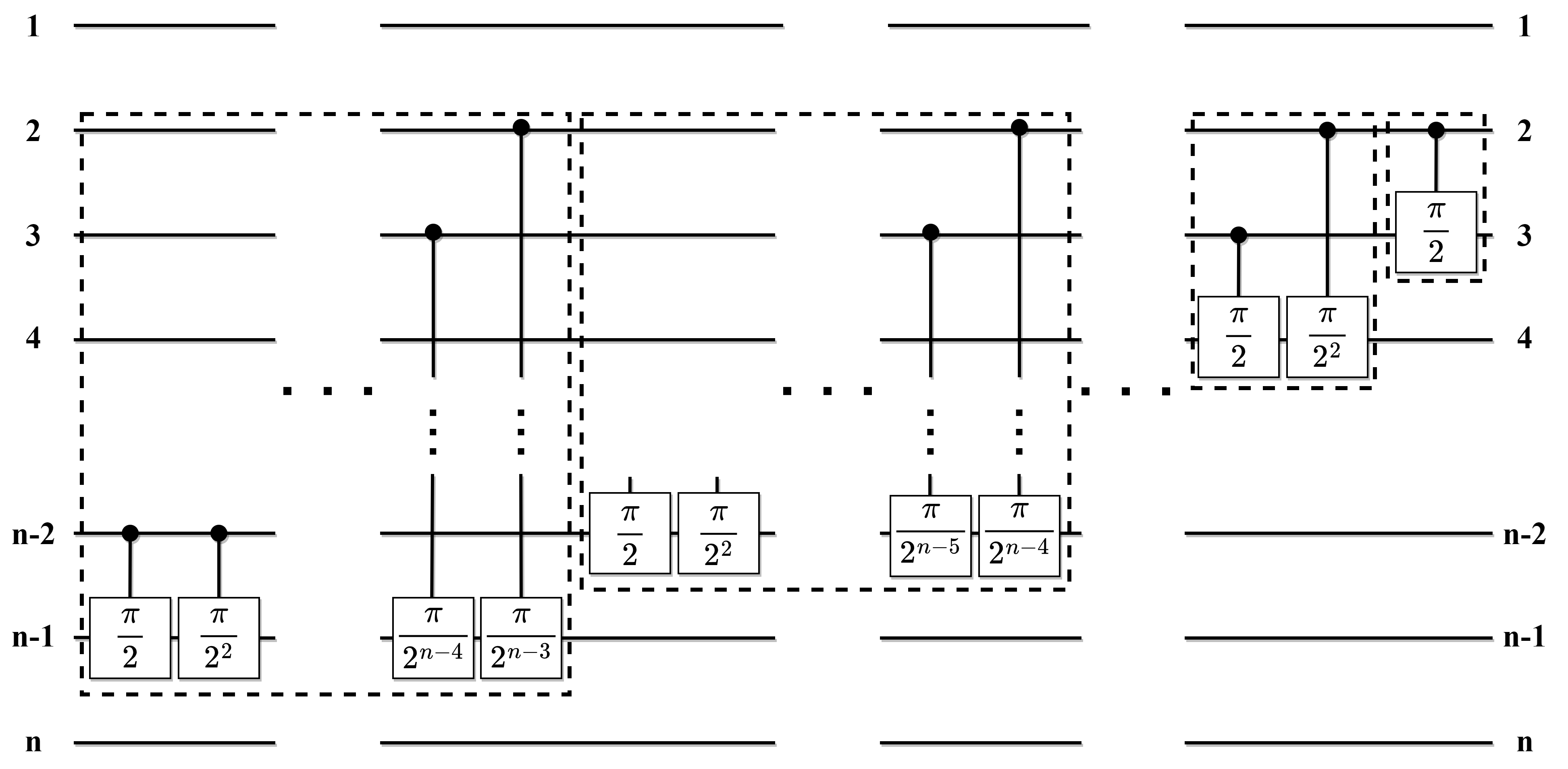}
			\vspace{0.3cm}
		\end{minipage}
		\begin{minipage}{1.05\linewidth}
			\leftline{\textbf{(b)}~$C_4^\prime:$}
			\vspace{0.3cm}
			\includegraphics[width=1\linewidth]{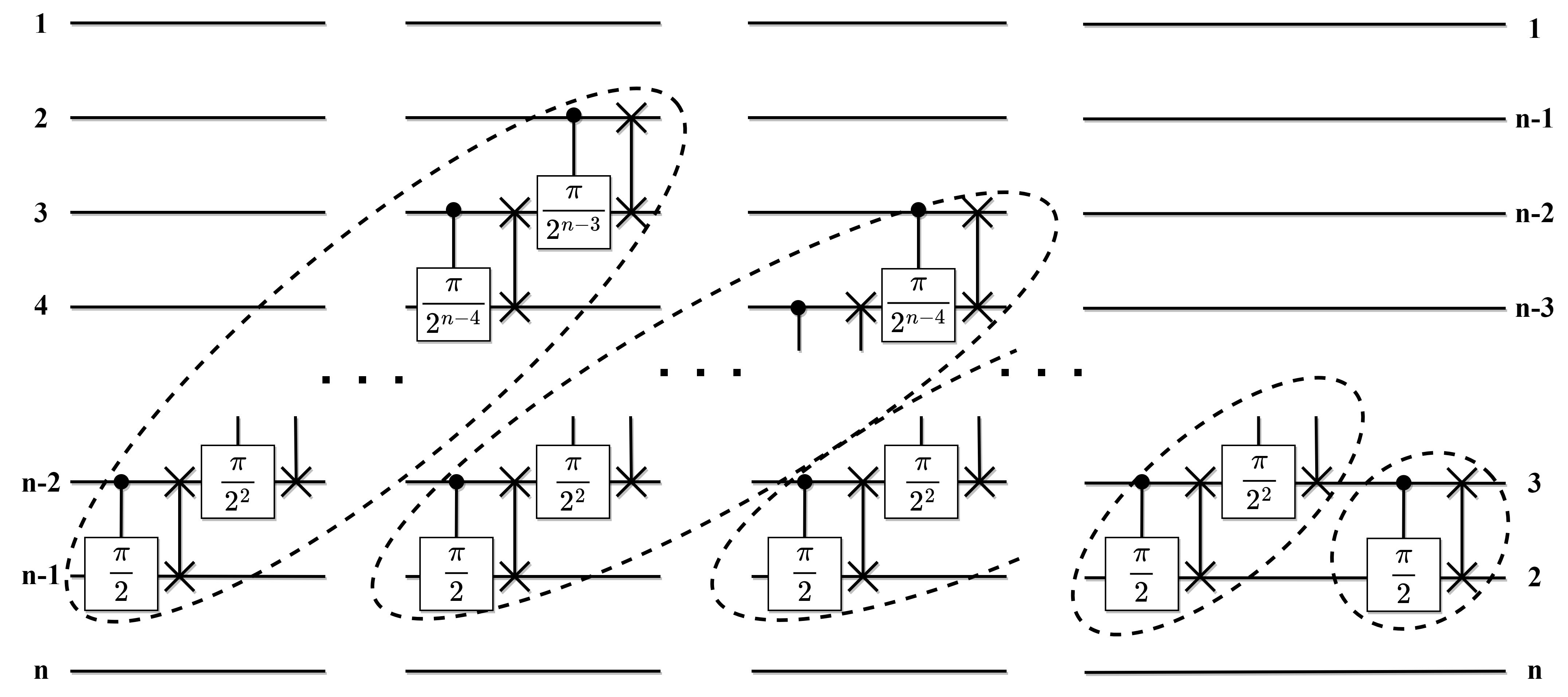}
		\end{minipage}
		\caption{\label{C_4_circuit}(a) An equivalent circuit for $C_4$ that only includes two-qubit controlled-$R_x(\theta)$ gates. (b) An equivalent circuit $C_4^\prime$ for $C_4$ that only includes local controlled-$R_x(\theta)$ and local swap gates. The qubit ordering is changed to $\{1,n-1,n-2,...,2,n\}$.}
	\end{figure}

\begin{figure}[htbp]
		\begin{minipage}{0.425\linewidth}
			\leftline{\textbf{(a)}~$C_5:$}
			\vspace{0.3cm}
			\includegraphics[width=1\linewidth]{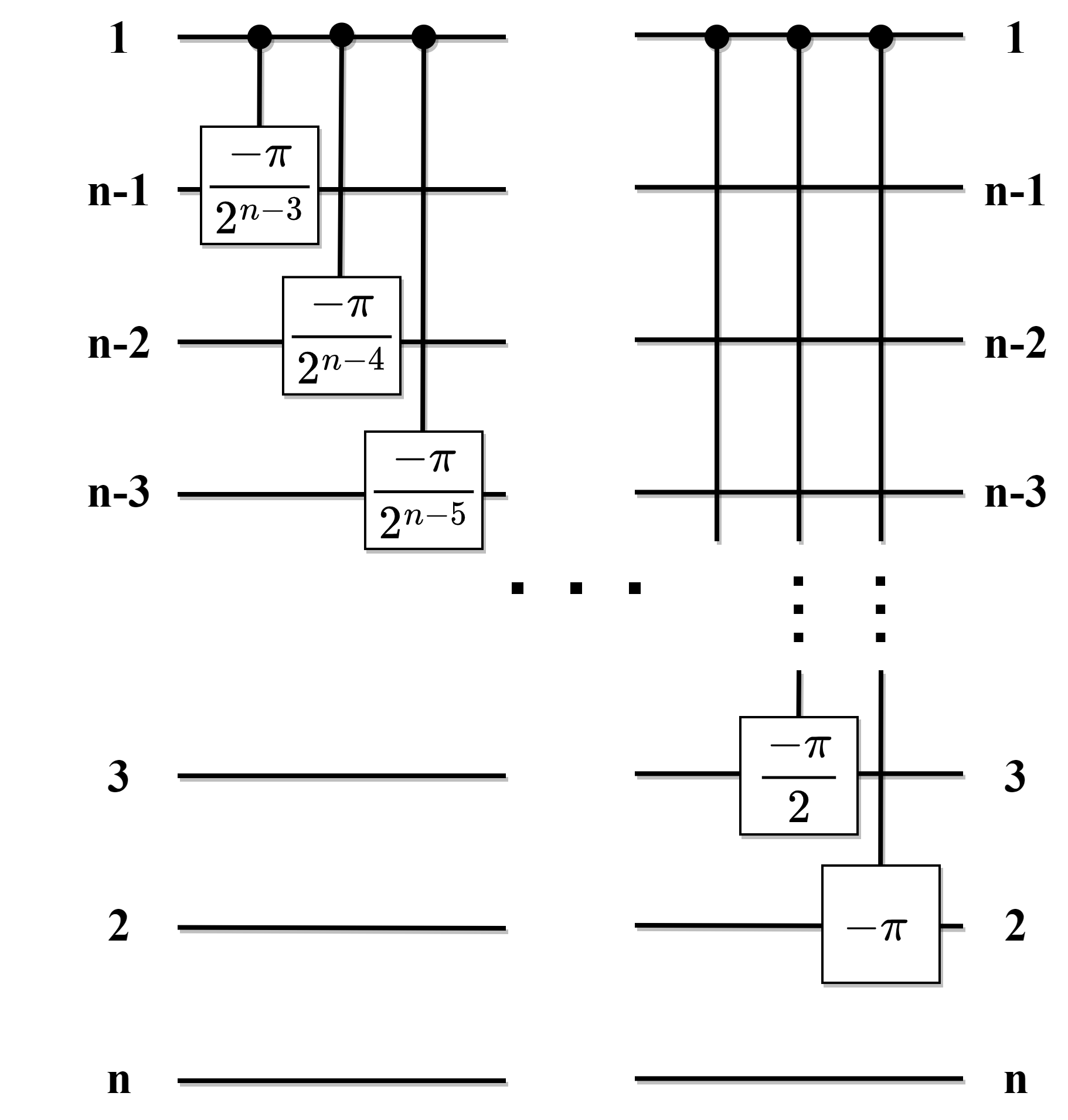}
		\end{minipage}
		\begin{minipage}{0.555\linewidth}
			\leftline{\textbf{(b)}~$C_5^\prime:$}
			\vspace{0.3cm}
			\includegraphics[width=1\linewidth]{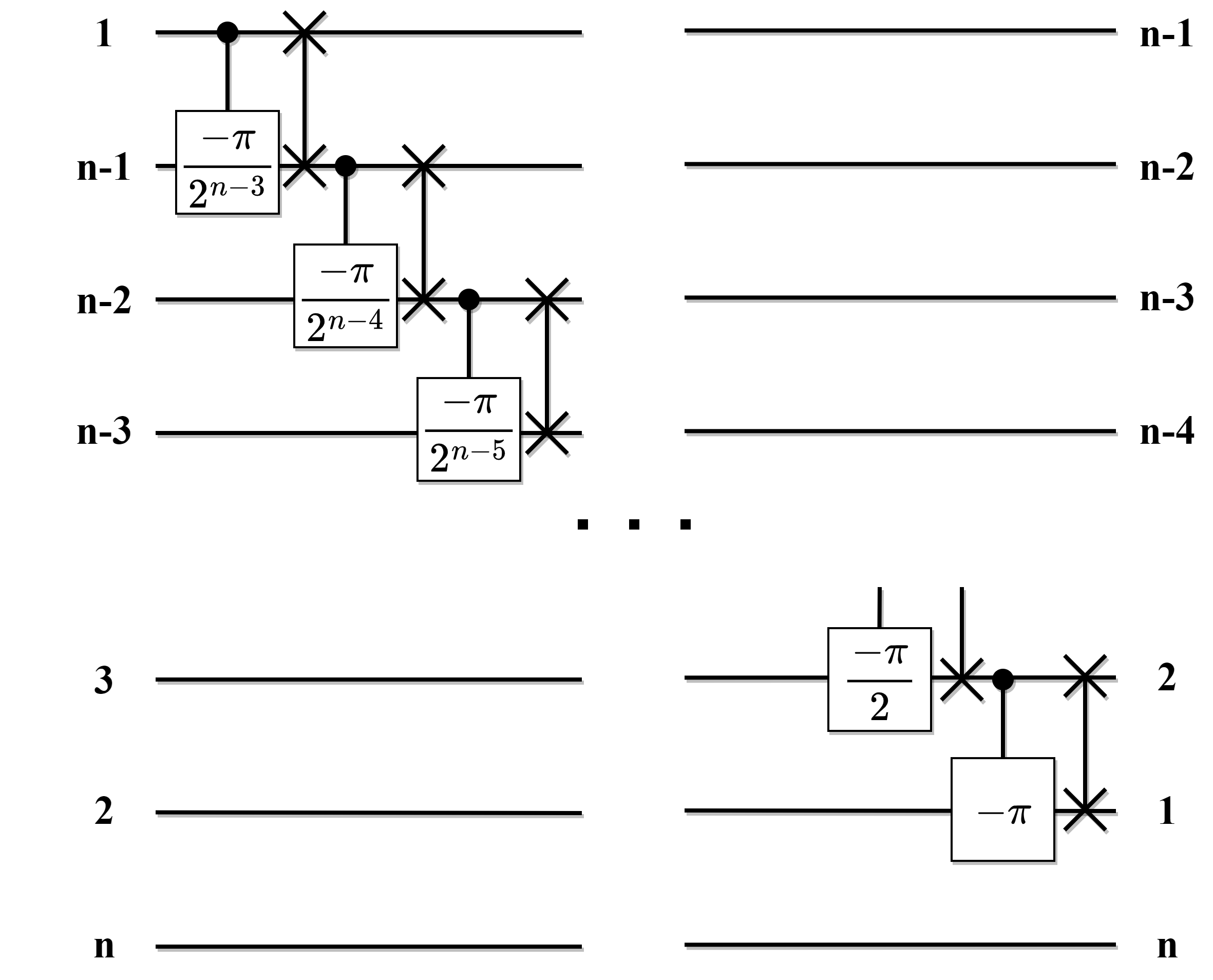}
		\end{minipage}
		\caption{\label{C_5_circuit}(a) An equivalent circuit for $C_5$ that only includes two-qubit controlled-$R_x(\theta)$ gates. (b) An equivalent circuit $C_5^\prime$ for $C_5$ that only includes local controlled-$R_x(\theta)$ and local swap gates. The qubit ordering is changed to $\{n-1,n-2,...,2,1,n\}$.}
	\end{figure}

\begin{figure}[htbp]
		\begin{minipage}{1\linewidth}
			\leftline{\textbf{(a)}~$C_6:$}
			\vspace{0.3cm}
			\includegraphics[width=1\linewidth]{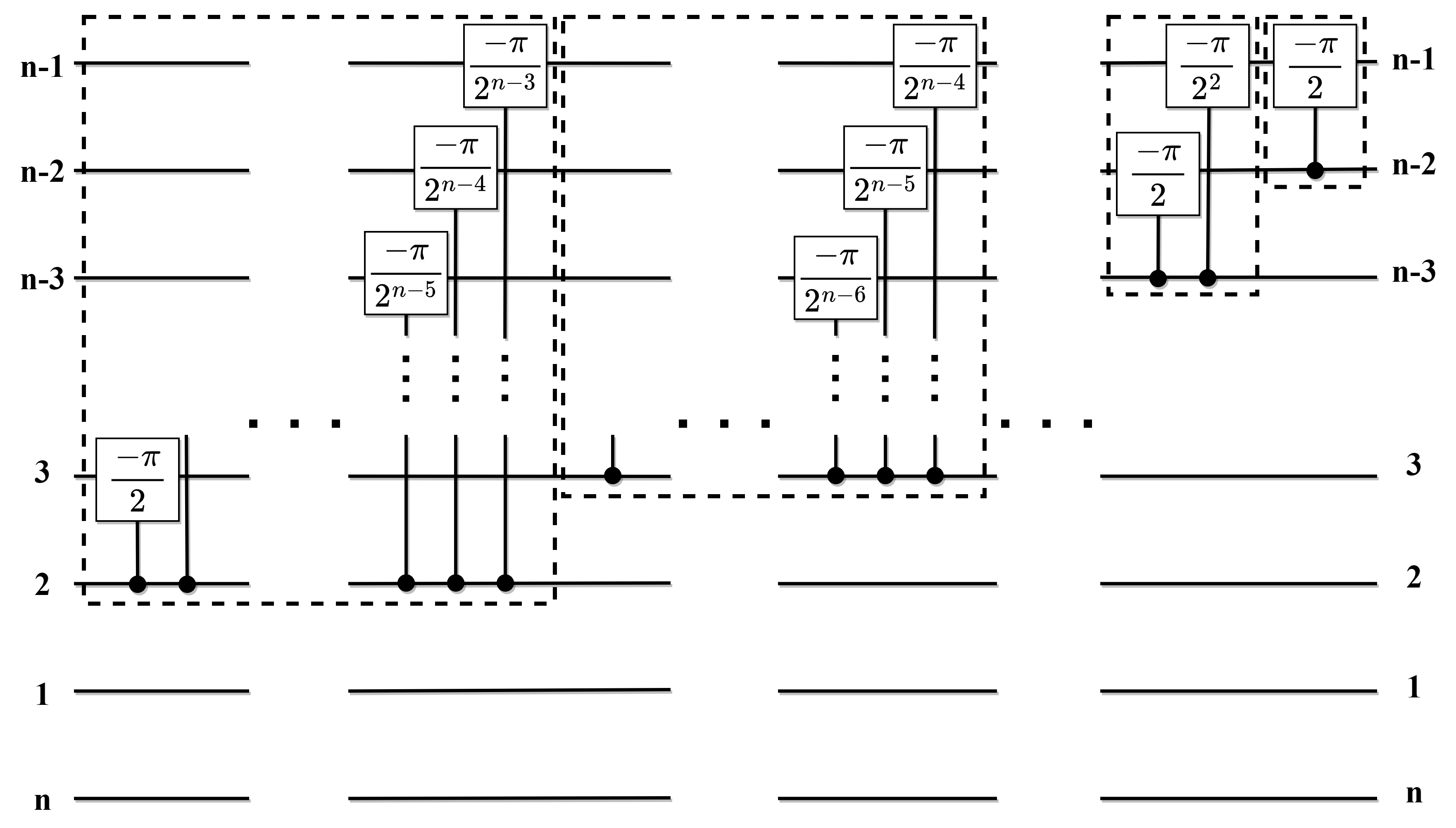}
			\vspace{0.3cm}
		\end{minipage}
		\begin{minipage}{1.05\linewidth}
			\leftline{\textbf{(b)}~$C_6^\prime:$}
			\vspace{-0.3cm}
			\includegraphics[width=1\linewidth]{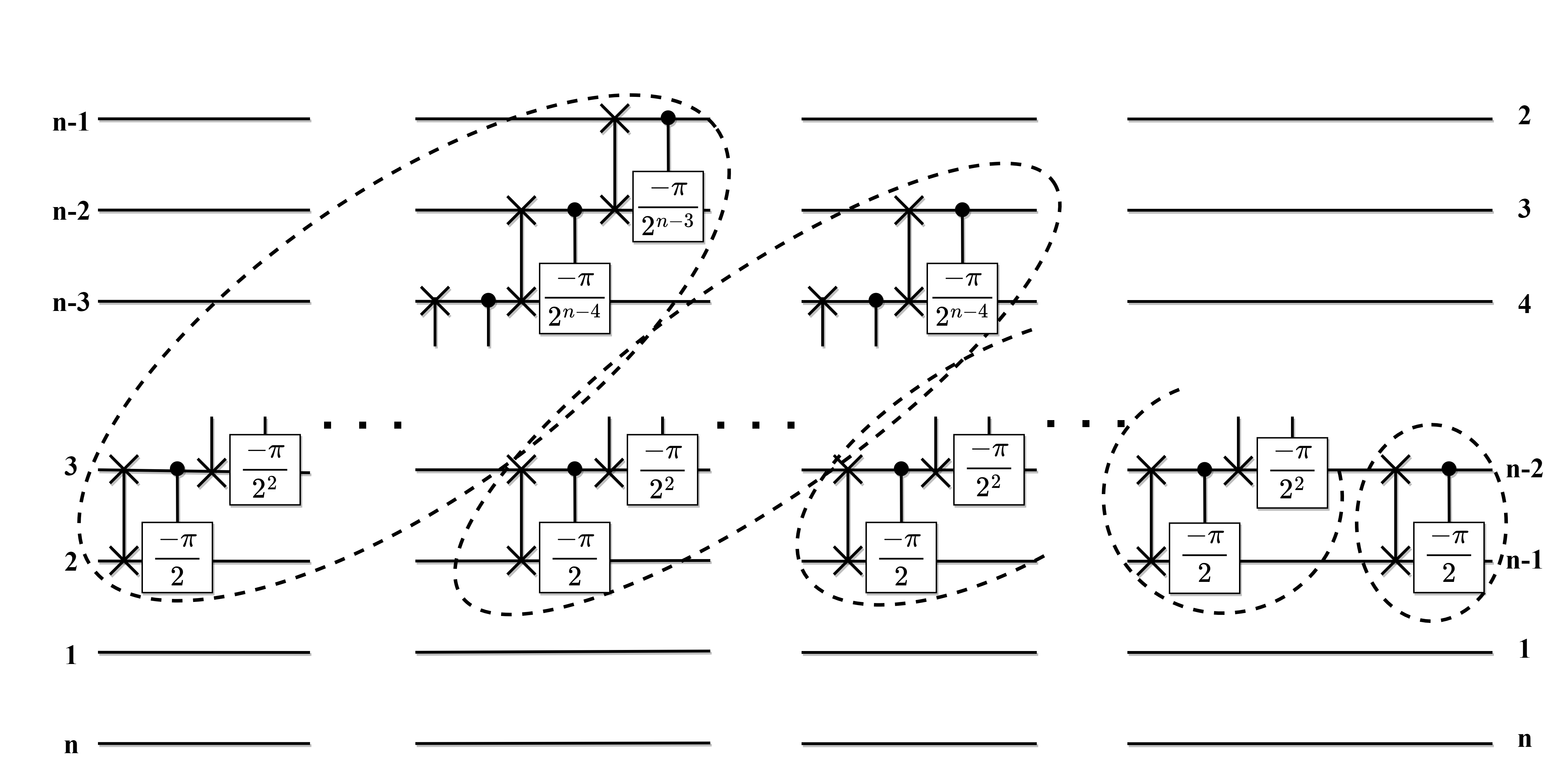}
		\end{minipage}
		\caption{\label{C_6_circuit}(a) An equivalent circuit for $C_6$ that only includes two-qubit controlled-$R_x(\theta)$ gates. (b) An equivalent circuit $C_6^\prime$ for $C_6$ that only includes local controlled-$R_x(\theta)$ and local swap gates. The qubit ordering is changed to $\{2,3,...,n-1,1,n\}$.}
	\end{figure}

\begin{figure}[htbp]
		\begin{minipage}{0.43\linewidth}
			\leftline{~$C_{6.5}^\prime:$}
			\vspace{0.3cm}
			\includegraphics[width=1\linewidth]{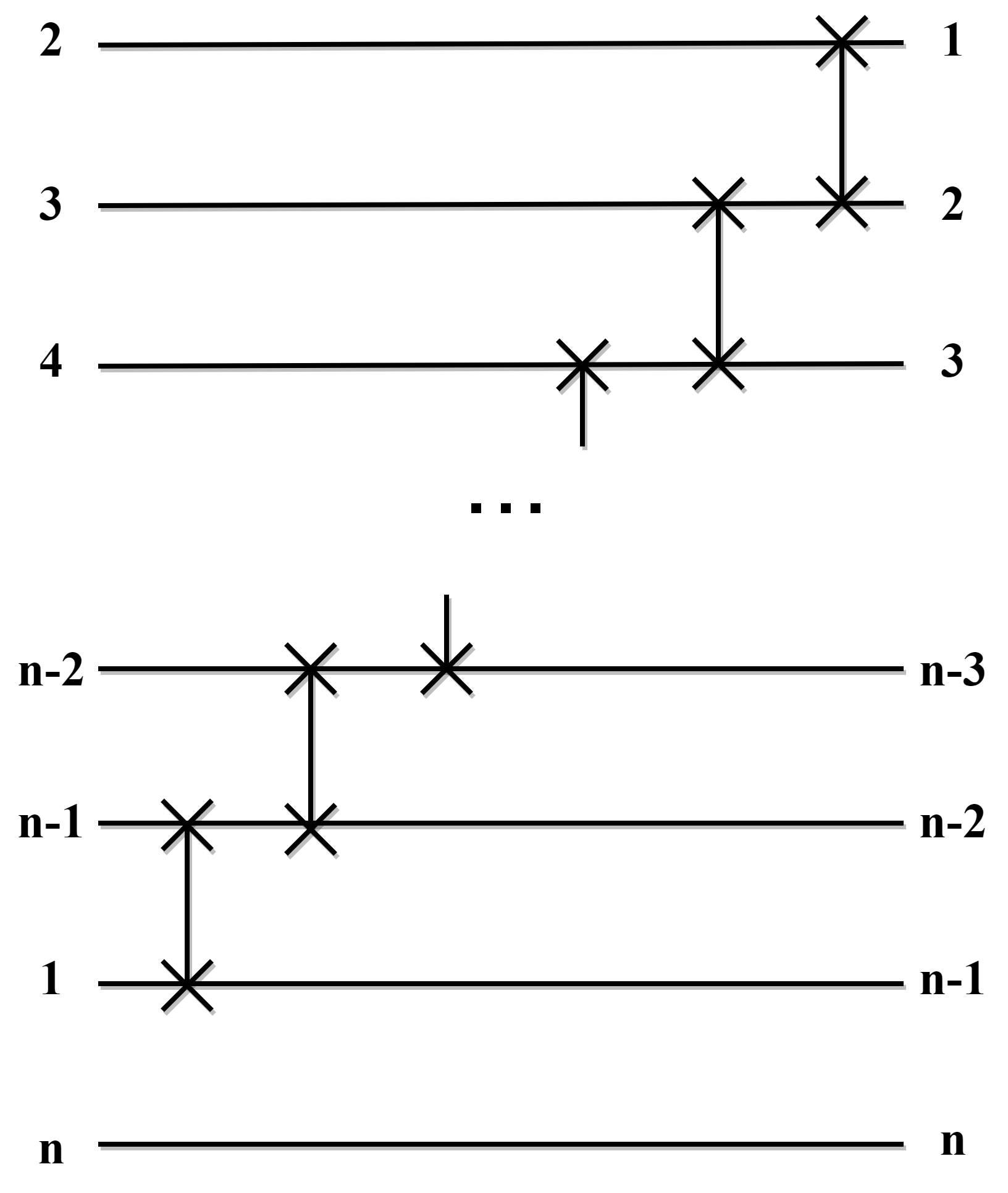}
		\end{minipage}
		\caption{\label{C_6_5}The circuit $C_{6.5}^\prime$ resets the qubit ordering $\{2,3,...,n-1,1,n\}$ to $\{1,2,3,...,n-1,n\}$.}
	\end{figure}

	Newly added parts $P_1$ and $P_2$ in Fig. \ref{Toffoli_circuit} are already the direct products of local gates. Part $C_1$ is reconstructed with two-qubit gates as shown in Fig. \ref{C_1_circuit}(a). Then it is proved that the circuit $C_1^\prime$ in Fig. \ref{C_1_circuit}(b) which only includes local gates is equivalent to $C_1$ but with qubit ordering changing: $\{1,2,3,...,n-1,n\}\to\{1,n,n-1,...,3,2\}$.
	With the new qubit ordering, we reconstruct $C_2$ as that only includes two-qubit and local gates as shown in Fig. \ref{C_2_circuit}(a) and (b) respectively.
	The qubit ordering is changed to $\{n,n-1,n-2,...,2,1\}$. We continue to reconstruct $C_3$ in Fig. \ref{C_3_circuit}(a) and (b) with two-qubit and local gates respectively.
	Now the qubit ordering is $\{2,3,...,n-1,n,1\}$. We apply a circuit $C_{3.5}^\prime$ to reset the qubit ordering to $\{1,2,3,...,n-1,n\}$ as shown in Fig. \ref{C_3_5}.
	Following the same procedure for decomposing $C_1$, $C_2$ and $C_3$, we decompose $C_4$, $C_5$ and $C_6$ as shown in Fig. \ref{C_4_circuit}, \ref{C_5_circuit} and \ref{C_6_circuit}. Finally a circuit $C_{6.5}^\prime$ is applied to reset the qubit ordering to be $\{1,2,3,...,n-1,n\}$ as shown in Fig. \ref{C_6_5}. In conclusion, the decomposition for $n$-qubit Toffoli gate is $P_1+C_1^\prime+C_2^\prime+C_3^\prime+C_{3.5}^\prime+P_2+C_4^\prime+C_5^\prime+C_6^\prime+C_{6.5}^\prime$.
	
	\section{Local gates in NMR systems}
	
	The local gates we used above are $\theta$-phase, swap and controlled-$R_x(\theta)$ gates. Swap gate can be decomposed into three CNOT gates. For instance, we will construct these local gates in NMR systems. Consider two qubits in NMR device. The time-evolution for qubit $1$ is $U_1(\theta_1)=e^{i\theta_1Z_1}$, for qubit $2$ is $U_2(\theta_2)=e^{i\theta_2Z_2}$ and for interaction between two qubits is $U_0(\theta_0)=e^{i\theta_0Z_1Z_2}$, where $Z_1$ and $Z_2$ are the Pauli-Z matrices for qubit $1$ and $2$ respectively \cite{Nielsen2002}. The $\theta$-phase gate for qubit $1$ can be obtained by directly applying $S(\theta)=\exp(i\theta/2)U_1(-\theta/2)$. Applying the control sequence $(I\otimes H)U_0(\theta_0)U_1(\theta_1)U_2(\theta_2)(I\otimes H)$ we obtain
	\begin{align}
		&\frac{1}{\sqrt{2}}
		\left(
		\begin{array}{cccc}
			1 & 1 & 0 & 0\\
			1 & -1 & 0 & 0\\
			0 & 0 & 1 & 1\\
			0 & 0 & 1 & -1
		\end{array}
		\right)
		\left(
		\begin{array}{cccc}
			e^{i\theta_0} & 0 & 0 & 0\\
			0 & e^{-i\theta_0} & 0 & 0\\
			0 & 0 & e^{-i\theta_0} & 0\\
			0 & 0 & 0 & e^{i\theta_0}
		\end{array}
		\right)\nonumber\\
		\times&\left(
		\begin{array}{cccc}
			e^{i\theta_1} & 0 & 0 & 0\\
			0 & e^{i\theta_1} & 0 & 0\\
			0 & 0 & e^{-i\theta_1} & 0\\
			0 & 0 & 0 & e^{-i\theta_1}
		\end{array}
		\right)
		\left(
		\begin{array}{cccc}
			e^{i\theta_2} & 0 & 0 & 0\\
			0 & e^{-i\theta_0} & 0 & 0\\
			0 & 0 & e^{i\theta_2} & 0\\
			0 & 0 & 0 & e^{-i\theta_2}
		\end{array}
		\right)\nonumber\\
		\times&\frac{1}{\sqrt{2}}
		\left(
		\begin{array}{cccc}
			1 & 1 & 0 & 0\\
			1 & -1 & 0 & 0\\
			0 & 0 & 1 & 1\\
			0 & 0 & 1 & -1
		\end{array}
		\right)\nonumber\\
		=&\frac{1}{2}\left(
		\begin{array}{cccc}
			\Delta_1+\Delta_2 & \Delta_1-\Delta_2 & 0 & 0\\
			\Delta_1-\Delta_2 & \Delta_1+\Delta_2 & 0 & 0\\
			0 & 0 & \Delta_3+\Delta_4 & \Delta_3-\Delta_4\\
			0 & 0 & \Delta_3-\Delta_4 & \Delta_3+\Delta_4
		\end{array}
		\right),\label{control_sequence}
	\end{align}
	where
	\begin{align}
		\Delta_1=&e^{i(\theta_0+\theta_1+\theta_2)},~~~~~\Delta_2=e^{i(-\theta_0+\theta_1-\theta_2)},\nonumber\\
		\Delta_3=&e^{i(-\theta_0-\theta_1+\theta_2)},~~~\Delta_4=e^{i(\theta_0-\theta_1-\theta_2)}.
	\end{align}
	Applying $\theta_0=\pi/4=-\theta_1=-\theta_2$, Eq. (\ref{control_sequence}) becomes to
	\begin{align}
		-\sqrt{i}\left(
		\begin{array}{cccc}
			1 & 0 & 0 & 0\\
			0 & 1 & 0 & 0\\
			0 & 0 & 0 & 1\\
			0 & 0 & 1 & 0
		\end{array}
		\right)
	\end{align}
	which is a CNOT gate. Applying $\theta_0=\theta/2$, $\theta_1=0$ and $\theta_2=-\theta/2$, Eq. (\ref{control_sequence}) becomes to
	\begin{align}
		\left(
		\begin{array}{cccc}
			1 & 0 & 0 & 0\\
			0 & 1 & 0 & 0\\
			0 & 0 & \cos\theta & -i\sin\theta\\
			0 & 0 & -i\sin\theta & \cos\theta
		\end{array}
		\right)
	\end{align}
	which is a controlled-$R_x(2\theta)$ gate.
	
	\section{Conclusion}
	Without deepening the circuit depth and leading in any ancilla, we modify the $n$-qubit controlled-$ZX$ gate in Ref. \cite{PhysRevA.87.062318} to be $n$-qubit Toffoli gate by only adding $2n-3$ $\theta$-phase gates. For practical application, we explicitly show the decomposition for $n$-qubit Toffoli gate with local gates. For instance, we construct these local gates in NMR systems.
	
	\begin{acknowledgments}
	We acknowledge the financial support in part by National Natural Science Foundation of China grant No.11974204 and No.12174215.
	\end{acknowledgments}
	

	\bibliography{refs}
	
\end{document}